\documentclass[twocolumn]{aastex631}
\begin{document}

\title{The Hubble PanCET program: Emission spectrum of hot Jupiter HAT-P-41b}

\author[0000-0002-3263-2251]{Guangwei Fu}
\affiliation{Department of Astronomy, University of Maryland, College Park, MD 20742, USA; guangweifu@gmail.com}

\author{David K. Sing}
\affiliation{Department of Physics and Astronomy, Johns Hopkins University, Baltimore, MD 21218, USA}

\author{Drake Deming}
\affiliation{Department of Astronomy, University of Maryland, College Park, MD 20742, USA; guangweifu@gmail.com}

\author[0000-0003-4552-9541]{Kyle Sheppard}
\affiliation{Department of Astronomy, University of Maryland, College Park, MD 20742, USA}

\author[0000-0003-4328-3867]{H.R. Wakeford}
\affil{School of Physics, University of Bristol, HH Wills Physics Laboratory, Tyndall Avenue, Bristol BS8 1TL, UK}

\author[0000-0001-5442-1300]{Thomas Mikal-Evans}
\affiliation{Department of Physics, and Kavli Institute for Astrophysics and Space Research, Massachusetts Institute of Technology, Cambridge, USA}

\author[0000-0003-4157-832X]{Munazza K. Alam}
\affiliation{Carnegie Earth \& Planets Laboratory, 5241 Broad Branch Rd NW, Washington, DC 20015, USA}

\author[0000-0002-2248-3838]{Leonardo A. Dos Santos}
\affil{Observatoire astronomique de l’Universit\'e de Gen\`eve, Chemin Pegasi 51, 1290 Versoix, Switzerland}
\affil{Space Telescope Science Institute, 3700 San Martin Drive, Baltimore, MD 21218, USA}

\author[0000-0003-3204-8183]{Mercedes L\'{o}pez-Morales}
\affiliation{Center for Astrophysics ${\rm \mid}$ Harvard {\rm \&} Smithsonian, 60 Garden Street, Cambridge, MA 02138, USA}

\begin{abstract}

We present the most complete emission spectrum for inflated hot Jupiter HAT-P-41b combining new HST WFC/G141 spectrum from the Hubble Panchromatic Comparative Exoplanet Treasury (PanCET) program with archival Spitzer eclipse observations. We found a near blackbody-like emission spectrum which is best fitted with an isothermal temperature-pressure (TP) profile that agrees well with the dayside heat redistribution scenario assuming zero Bond albedo. The non-inverted TP profile is consistent with the non-detection of NUV/optical absorbers in the transit spectra. We do not find any evidence for significant H$^-$ opacity nor a metal-rich atmosphere. HAT-P-41b is an ideal target that sits in the transitioning parameter space between hot and ultra-hot Jupiters, and future JWST observations will help us to better constrain the thermal structure and chemical composition.

\end{abstract}
\nopagebreak

\section{Introduction}

Emission spectroscopy of exoplanets allows us to probe into the deeper layers (1-10 bar) of the dayside atmosphere compared to transmission spectroscopy which measures the upper layers ($\sim$1mbar) of the planetary limbs. The difference in the radiative transfer path geometry means the emission spectrum is much more sensitive to the vertical thermal structure of the planet. Depending on the atmospheric chemical composition and the temperature-pressure profile, molecular absorption or emission features can be imprinted onto the emission spectrum. Notably, water features at 1.4 $\mu m$ \citep{fu_statistical_2017} has been the most robustly detected in both absorption \citep{kreidberg_precise_2014} and emission \citep{fu_strong_2022,evans_ultrahot_2017}. The presence of H- continuum opacity was inferred in multiple ultra-hot Jupiters \citep{arcangeli_climate_2019, fu_hubble_2021-1}. In the longer wavelength, the CO/CO$_2$ features were indicated by the deviations of Spitzer photometric points in 3.6 $\mu m$ and 4.5 $\mu m$ channels from the blackbody approximation \citep{garhart_statistical_2020, fu_hubble_2021}.

HAT-P-41b is an inflated hot Jupiter (R=$1.685^{+0.076}_{-0.051}$R$_{Jup}$ M=$0.8^{+0.1}_{-0.1}$M$_{Jup}$ T$_{eq}$=$1940^{+38}_{-38}$K) \citep{hartman_hat-p-39bhat-p-41b_2012} with detailed atmospheric characterization via transmission spectroscopy \citep{sheppard_hubble_2021, wakeford_into_2020, lewis_into_2020}. The planet shows a high metallicity \citep{wakeford_into_2020, sheppard_hubble_2021} atmosphere with increased H- opacity abundance \citep{lewis_into_2020}. With a dayside temperature of $\sim$2300K, HAT-P-41b sits between hot and ultra-hot Jupiters where physical processes such as molecular dissociation and H- opacity are starting to become important \citep{parmentier_thermal_2018}. HAT-P-41b allows us to pinpoint the transitional parameter space of various atmospheric processes which makes it a valuable data point in our understanding of hot Jupiter atmospheres on a population level \citep{mansfield_unique_2021, baxter_evidence_2021}. Here we present the 1.1 to 4.5 $\mu m$ emission spectrum of HAT-P-41b with combined data from HST/WFC3 G141 and $Spitzer$ channel 1 and 2.

\begin{figure*}
\centering
  \includegraphics[width=\textwidth,keepaspectratio]{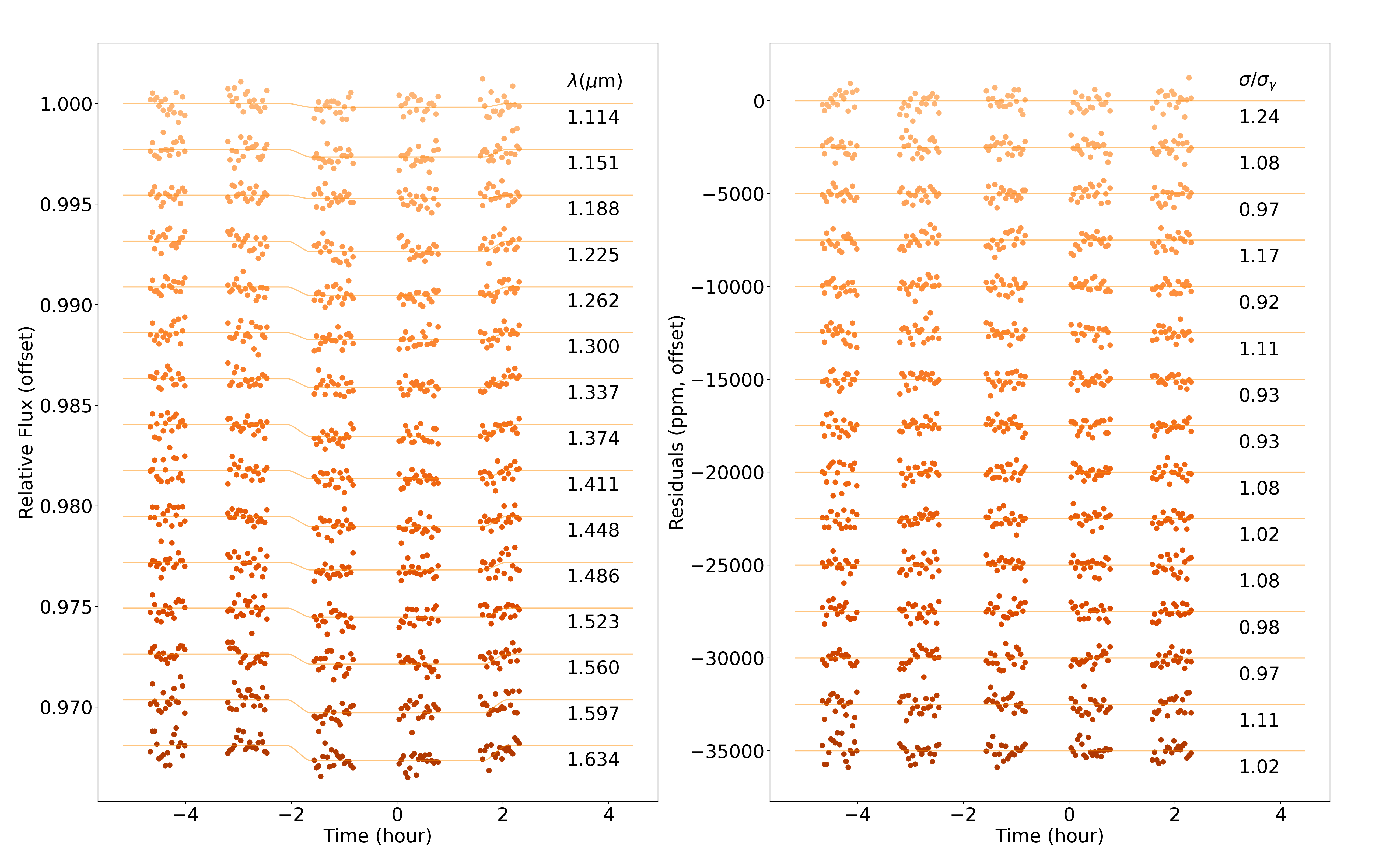}
  \caption{HST WFC3 G141 spectral bin transit lightcurves after ramp-effect correction using \texttt{RECTE} (left) and corresponding residuals (right) with their relative ratios to the photon-limit noise ($\sigma/\sigma_\gamma$) levels.}
  \label{fig:WFC3}
\end{figure*}

\begin{figure}
\centering
  \includegraphics[width=0.5\textwidth,keepaspectratio]{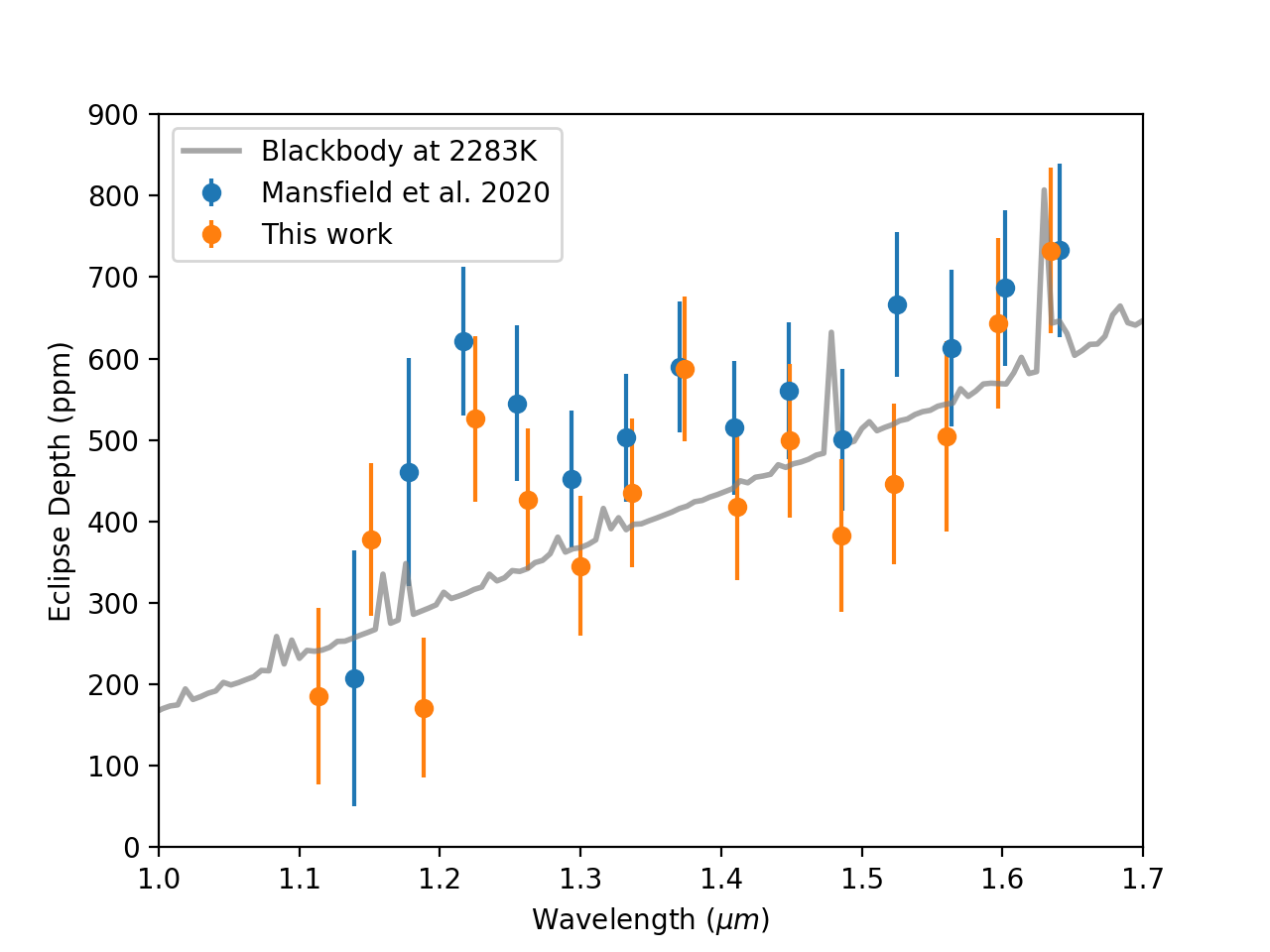}
  \caption{The WFC3/G141 emission spectrum is in excellent agreement with \cite{mansfield_unique_2021}.}
  \label{fig:mansfield}
\end{figure}

\section{Observations}

The $Hubble$ WFC3/G141 dataset was observed as a part of the Panchromatic Exoplanet Treasury Survey (PanCET GO 14767, PIs: Sing $\&$ Lopez-Morales) on 2016-10-09 and the $Spitzer$ dataset was published in \citep{garhart_statistical_2020}. The WFC3/G141 eclipse observation was taken in spatial scan mode for five consecutive HST orbits. Each frame was taken with the 256 $\times$ 256 pixel subarray in the SPARS10 and NSAMP=12 settings. The forward scanning rate is 0.065 arcsec $s^{-1}$ and the exposure time is 81 seconds.

\section{HST WFC3 data reduction}

All the orbital parameters for the WFC3/G141 data reduction have been fixed to the same values used in \citep{sheppard_hubble_2021} which are identical to what were used in the \citep{garhart_statistical_2020} $Spitzer$ analysis.

The data reduction process starts with applying the standard flat field, background subtraction and bad pixels, cosmic rays removal on the ima frames \citep{fu_hubble_2021-1}. Then we extract the non-destructive reads from each frame \citep{deming_infrared_2013}. There is a spatially resolved companion star \citep{sheppard_hubble_2021} located 3.6 arcseconds away. Due to the large spatial separation, the non-destructive reads of spatial scan from the two stars do not overlap which allows for clean removal of the companion star spectra. The companion star removed reads are then combined to the complete spatial scan frames. Next we summed each frame in the vertical direction for the 1D spectrum and normalized it by its own median flux to calculate the wavelength shifts. We then used $scipy.interpolate.interp1d$ function to interpolate the 1D spectrum in the wavelength direction and calculated the relative sub-pixel level shifts of each frame. The largest shift between any two frames is under 0.1 pixel which does not induce any excessive systematics \citep{stevenson_analyzing_2019}. Wavelength shifts corrected non-normalized frames were then summed in all wavelength channels to form the whitelight eclipse lightcurve. The lightcurve is then fitted with a combination of \texttt{BATMAN} \citep{kreidberg_batman_2015} with the \texttt{RECTE} charge trapping systematics model \citep{zhou_physical_2017}, 2nd order polynomial of the HST orbital phase and the wavelength shifts. Each wavelength channel is then fitted using the same method but with the mid-transit time fixed to the best-fit whitelight value. The best-fit lightcurves of each wavelength channel and corresponding residuals are shown in Figure \ref{fig:WFC3}.

\begin{figure*}
\centering
  \includegraphics[width=\textwidth,keepaspectratio]{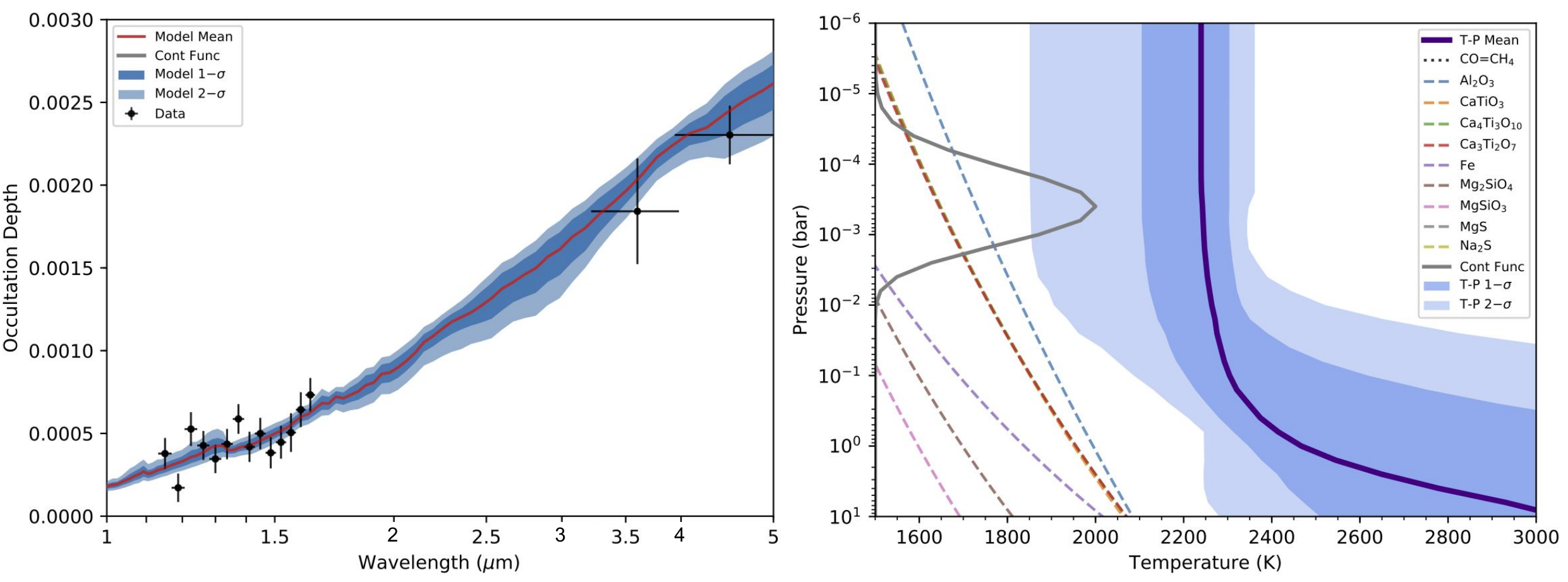}
  \caption{ATMO retrieval of HAT-P-41b emission spectrum. The featureless black-like spectrum (left) is best fitted with an isothermal TP profile (right).}
  \label{fig:atmo_retrieval}
\end{figure*}

\begin{figure}
\centering
  \includegraphics[width=0.45\textwidth,keepaspectratio]{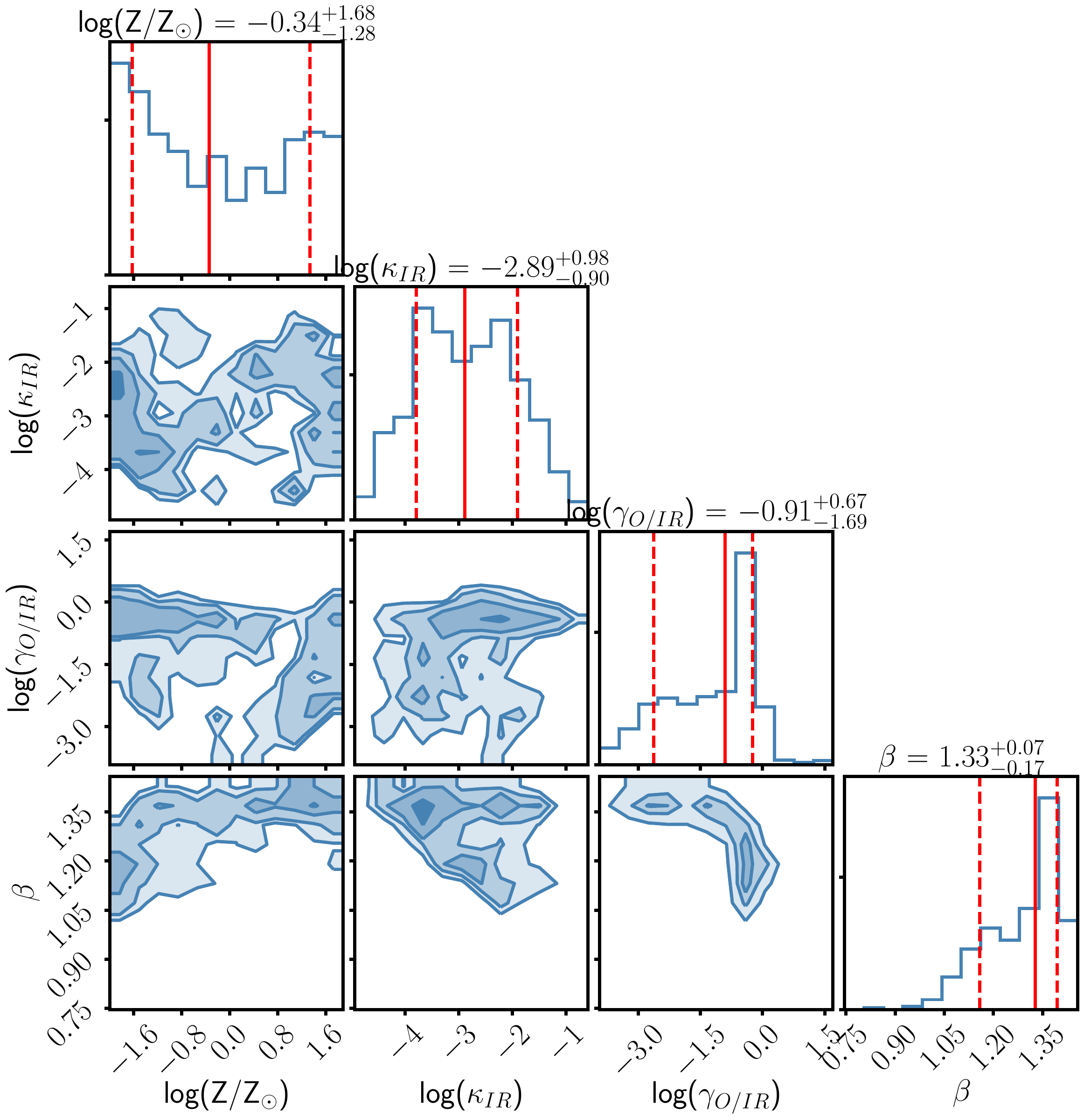}
  \caption{Posterior distribution of the ATMO retrieval. The metallicity is mostly unconstrained due to a lack of features within the emission spectrum.}
  \label{fig:atmo_corner}
\end{figure}

\begin{figure*}
\centering
  \includegraphics[width=\textwidth,keepaspectratio]{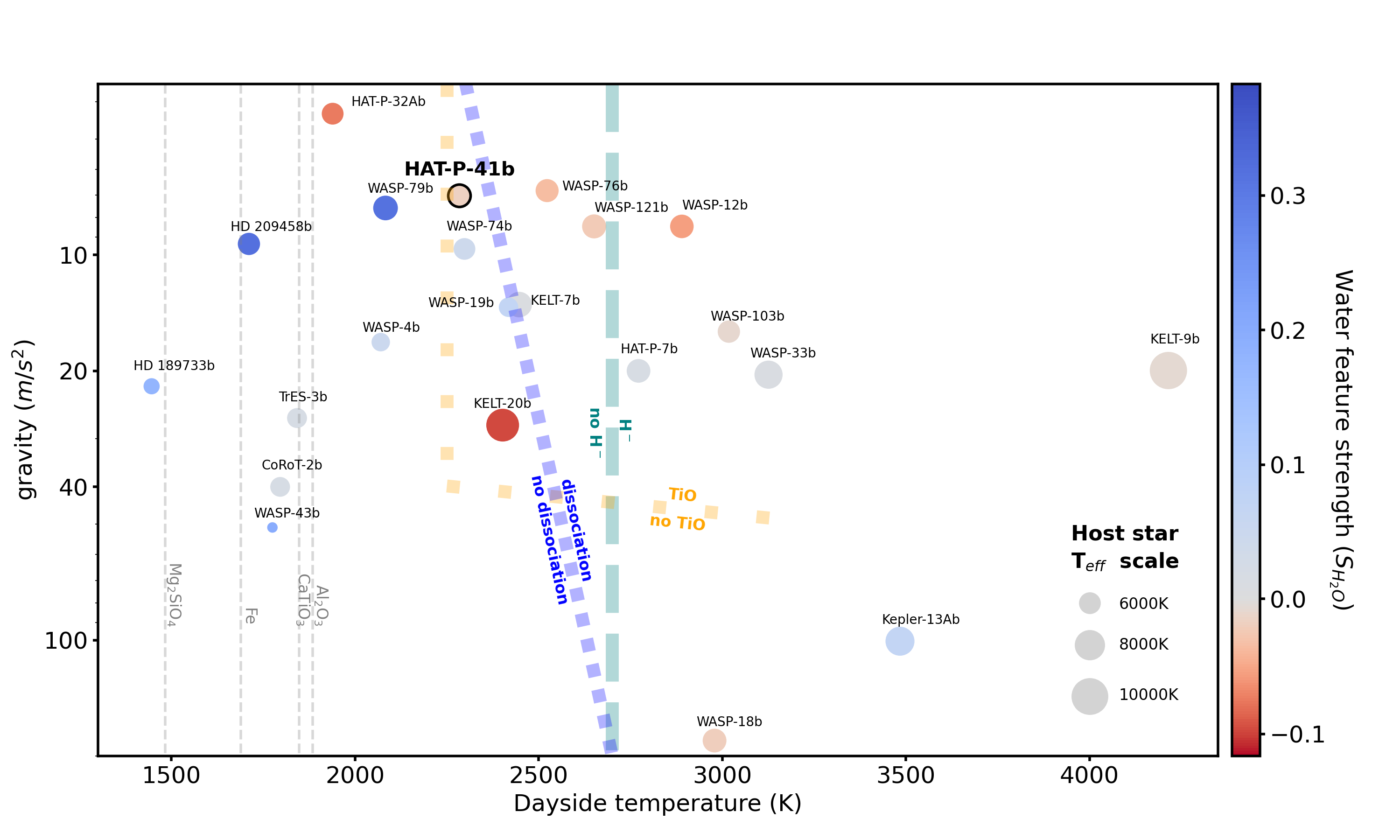}
  \caption{HAT-P-41b WFC3/G141 emission spectrum compared with other hot Jupiters. The water feature strength index (S$_{H_2O}$) \citep{mansfield_unique_2021} measures the 1.4$\mu m$ water band feature size relative to the blackbody model based on the out-of-band spectrum. Centered at zero being the blackbody-like featureless spectrum. Increasing negative values (redder color) represent stronger emission, features and increasing positive values (bluer color) represent stronger absorption features. The circle size scales with the host star temperature (T$_{eff}$). The grey vertical dash lines are condensation lines for various metals at 100 mbar assuming solar metallicity. We see mostly featureless spectra at higher dayside temperatures ($>$2700K) where H$^-$ continuum opacity expects to dominate \citep{parmentier_thermal_2018}. At the lower temperature side, we see mostly water absorption features where the atmosphere is too cool for gaseous heavy metal absorbers such as TiO to form and drive thermal inversion. HAT-P-41b sits at this in-between transitional region where planets can have absorption, emission or featureless spectra. Further observations of HAT-P-41b in the longer wavelength with JWST will help us to precisely determine its thermal structure and understand what causes thermal inversion in hot Jupiters atmospheres.}
  \label{fig:compare}
\end{figure*}

\section{Retrieval analysis}

We performed the retrieval analysis on the reduced HAT-P-41b emission spectrum (Table. \ref{eclipse_spectrum}) with \texttt{ATMO} \citep{amundsen_accuracy_2014, drummond_effects_2016, goyal_library_2018, tremblin_fingering_2015, tremblin_cloudless_2016, wakeford_hat-p-26b_2017}. The retrieval setup is equilibrium chemistry with a fixed solar C/O ratio. There is a total of 4 free fitting parameters (Table. \ref{eclipse_priors}) including metallicity (Z) and three parameters $\kappa_{IR}$, $\gamma_{O/IR}$ and $\beta$ for the temperature-pressure (TP) profile as defined in \cite{line_systematic_2013}. We fixed the C/O ratio to the solar value due to the lack of constrain this dataset has on the carbon-bearing species. Also, the transit spectra retrieval \citep{sheppard_hubble_2021, lewis_into_2020} results were all consistent with the solar C/O values.

\begin{center}
\begin{deluxetable}{ccc}
\tablecaption{\textbf{ATMO eclipse retrieval priors $\&$ posteriors}}
\tablehead{\colhead{Parameter} & \colhead{Priors} & \colhead{Posteriors}}
\startdata
\hline\hline
log(Z/Z$_{\Sun}$)	&	$\mathcal{U}$(-2.8, 2.8)	&	$-0.342^{+1.677}_{-1.277}$	\\
log(K$_{IR}$)	&	$\mathcal{U}$(-5, -0.5)	&	$-2.887^{+0.982}_{-0.907}$	\\
log($\gamma$/IR)	&	$\mathcal{U}$(-4, 1.5)	&	$-0.907^{+0.672}_{-1.716}$	\\
beta	&	$\mathcal{U}$(0, 2)	&	$1.326^{+0.068}_{-0.168}$	\\
\hline
\enddata
\end{deluxetable}
\end{center}
\label{eclipse_priors}

The best-fit model to the emission spectrum from the \texttt{ATMO} retrieval (Figure. \ref{fig:atmo_retrieval}) gives a $\chi^2_{\nu}$ of 1.24 (4 degrees of freedom) with a near-isothermal TP profile. The best-fit blackbody temperature using the PHOENIX stellar model \citep{husser_new_2013} grid (logg = 4.5 and logZ = 0) interpolated to T$_{eff}$ = 6390 K is 2283$\pm$64K with a $\chi^2_{\nu}$ of 1.23 (1 degree of freedom). Based on the formalism described in \citep{cowan_statistics_2011}, this dayside temperature would suggest a circulation efficiency $\varepsilon$ of 0.44 assuming zero Bond albedo and the predicted nightside temperature would be 1572K. $\varepsilon$=0 represents the no-circulation limit where the nightside temperature is 0K and $\varepsilon$=1 represents the full-circulation scenario where nightside temperature is the same as the dayside. The measured dayside temperature is consistent with the dayside-only heat redistribution scenario. The retrieved metallicity is consistent with the solar value (Figure. \ref{fig:atmo_corner}). While \cite{sheppard_hubble_2021} retrieved metallicity $\sim$1$\sigma$ higher than solar, \cite{lewis_into_2020} reported values consistent with solar metallicity. Considering the large uncertainties on metallicity from all three studies, the current datasets can not well constrain the models.

\cite{lewis_into_2020} found H- abundance several orders of magnitude larger than equilibrium chemistry is needed to best fit the transmission spectrum. We did not find evidence for significant H$^{-}$ opacity from the emission spectrum and we believe this could be due to: (1) The blackbody-like emission spectrum and near-isothermal TP profile make the abundances largely unconstrained. (2) The H- constrain in \cite{lewis_into_2020} comes from a combination of WFC3 G280 and G141 data. \cite{wakeford_into_2020} did not find evidence for H- solely based on the G280 optical dataset. Therefore any uncorrected offsets between G280 and G141 spectra can potentially lead to high H- abundance. (3) \cite{sheppard_hubble_2021} did not retrieve significant H- abundance with both G280 and G141 datasets using a different retrieval code. So it could also be due to different forward model assumptions.

Heavy metals such as Fe I, Fe II, and molecule like TiO are considered major optical absorbers that can induce temperature inversion in hot Jupiter atmospheres \citep{fortney_unified_2008, lothringer_uv_2020}, and they have been detected in the optical transit spectrum with HST/STIS G430L and G750L \citep{fu_hubble_2021-1}. There are no significant NUV heavy metals nor optical TiO absorption detected in both STIS \citep{sheppard_hubble_2021} and UVIS \citep{wakeford_into_2020, lewis_into_2020} transmission spectra for HAT-P-41b. The non-inverted TP profile retrieved in this HAT-P-41b emission spectrum is consistent with the non-detection of NUV/optical absorbers in the transit spectra.

\section{Compared with other hot Jupiter atmospheres}

To quantify the WFC3/G141 emission spectra from various hot Jupiters and compare HAT-P-41b to them, we took the 1.4$\mu m$ water feature strength index (S$_{H_2O}$) reported in \cite{mansfield_unique_2021} and plotted them in the dayside temperature versus gravity parameter space (Figure \ref{fig:compare}) first introduced in \cite{parmentier_thermal_2018}. The important physical transitions including TiO, H$^-$, and molecular dissociation are also the same as shown in \cite{parmentier_thermal_2018}. The larger negative S$_{H_2O}$ values (redder) represent stronger water emission features while the larger positive values (bluer) indicate more prominent water absorption features. The index is centered at zero for a blackbody-like featureless spectrum. In addition, we also scale the size of each point based on the host star temperature (T$_{eff}$) to reflect the effect of increased FUV/UV radiations from earlier type stars \citep{lothringer_influence_2019, fu_strong_2022}.

We have only seen strong water absorption features among cooler hot Jupiter (T$_{day}<$2200K) atmospheres \citep{kreidberg_precise_2014, line_no_2016} driven by their decreasing TP profiles. This is due to heavy metal absorbers condensing out of the atmosphere at low temperatures. On the other hand, when the atmosphere becomes too hot (T$_{day}>$2700K), water molecule dissociates and H$^-$ continuum opacity starts to dominate. As a result, we have only observed featureless emission spectra among the hottest planets. In between the two regions ($\sim$2200K to 2700K) we see a transitional parameter space where atmospheres can be inverted \citep{fu_strong_2022, evans_ultrahot_2017} or isothermal \citep{fu_hubble_2021, mansfield_unique_2021}. HAT-P-41b sits in the middle of this transitional space. Its mostly featureless WFC3/G141 emission spectrum is similar to that of WASP-76b, WASP-74b, WASP-19b and KELT-7b. However, it is different from WASP-121b and KELT-20b where we saw evidence for water emissions. Although previous models have indicated multiple important physical transitions \citep{parmentier_thermal_2018} happening at this temperature and gravity range, it is not yet well understood what exact physical processes drive the different thermal structures and emission spectra of these planets. In addition, host star type could be another important determining factor since strong FUV/UV radiations from the host star can strengthen atmospheric thermal inversion \citep{fu_strong_2022, lothringer_influence_2019} as they deposit significant energy into the upper layers of the planetary atmospheres. This is supported by the large emission features saw in KELT-20b which orbits an A-type star. The lack of FUV/UV flux from F-type host star of HAT-P-41b could be another cause for its isothermal atmosphere.

\section{Conclusion}

We present the most complete emission spectrum for inflated hot Jupiter HAT-P-41b. The spectrum is close to blackbody-like with no significant molecular absorption or emission features. The best-fit \texttt{ATMO} model shows an isothermal TP profile agreeing with the dayside heat redistribution scenario and a metallicity value consistent with the solar value. The non-inverted TP profile is consistent with the non-detection of NUV/optical absorbers in the transit spectra. Significant H$^-$ opacity suggested in \cite{lewis_into_2020} is not required in the model to adequately fit the emission spectrum. We also do not retrieve a metal-rich atmosphere as indicated in \citep{sheppard_hubble_2021}. However, the emission spectrum does not well constrain the atmosphere metallicity due to the limited wavelength coverage. The featureless emission spectrum of HAT-P-41b indicates planets with a dayside temperature around 2300K may have relatively isothermal TP profiles in the absence of heavy metal absorbers and strong host star FUV/UV radiations to drive thermal inversion. The comparison of HAT-P-41b to other similar hot Jupiters paints a murky picture of how atmospheric physical properties transition from cooler to hotter planets. The combined effect of surface gravity, thermal dissociation, H$^-$ opacity, heavy metal absorbers, and host star type is yet to be disentangled. Future observations of more similar planets and follow-up JWST infrared measurements of HAT-P-41b will be the key to solving the mystery of hot Jupiter atmospheres.

\startlongtable
\begin{center}
\begin{deluxetable*}{cccc}
\tablecaption{\textbf{HAT-P-41b eclipse spectrum}}
\tablehead{\colhead{Wavelength midpoint ($\mu$m)} & \colhead{Bin width ($\mu$m)} & \colhead{Occultation Depth (ppm)} & \colhead{Uncertainty (ppm)}}
\startdata
1.1137	&	0.0186	&	186	&	108	\\
1.1509	&	0.0186	&	378	&	94	\\
1.1881	&	0.0186	&	171	&	86	\\
1.2253	&	0.0186	&	526	&	102	\\
1.2625	&	0.0186	&	427	&	87	\\
1.2997	&	0.0186	&	346	&	86	\\
1.3369	&	0.0186	&	435	&	91	\\
1.3741	&	0.0186	&	587	&	89	\\
1.4113	&	0.0186	&	419	&	91	\\
1.4485	&	0.0186	&	499	&	95	\\
1.4857	&	0.0186	&	383	&	94	\\
1.5229	&	0.0186	&	446	&	99	\\
1.5601	&	0.0186	&	505	&	117	\\
1.5973	&	0.0186	&	643	&	105	\\
1.6345	&	0.0186	&	733	&	102	\\
3.6000	&	0.3800	&	1842	&	319	\\
4.5000	&	0.5600	&	2303	&	177	\\
\hline
\enddata
\label{eclipse_spectrum}
\end{deluxetable*}
\end{center}

\bibliography{references}

\clearpage

\end{document}